\documentstyle[prd,aps,epsf,amsfonts,amssymb,amsmath]{revtex}
\newcommand{\be}{\begin{equation}}\newcommand{\ee}{\end{equation}}
\newcommand{\bea}{\begin{eqnarray}}\newcommand{\eea}{\end{eqnarray}}
\newcommand{\brr}{\begin{array}}\newcommand{\err}{\end{array}}
\newcommand{\bit}{\begin{itemize}}\newcommand{\eit}{\end{itemize}}
\newcommand{\ben}{\begin{enumerate}}\newcommand{\een}{\end{enumerate}}

\def\noi{\noindent}

\def\1{{_{1}}}\def\2{{_{2}}}
\begin{document}

\title{On observability of R\'{e}nyi's entropy}

\vspace{5mm}

\author{Petr Jizba${}^{}$ and
Toshihico Arimitsu${}^{}$}

\address{ $ $\\[2mm]
${}^{}$ Institute of Physics, University of Tsukuba, Ibaraki
305-8571, Japan
\\ [2mm] E-mails: petr@cm.ph.tsukuba.ac.jp, arimitsu@cm.ph.tsukuba.ac.jp
 }

\maketitle

\vspace{5mm}
\begin{center}
{\small \bf Abstract}
\end{center}
\vspace{-2mm}
\begin{abstract}
\hspace{-5.5mm} Despite recent claims we argue that R\'{e}nyi's
entropy is an observable quantity. It is shown that, contrary to
popular belief, the reported domain of instability for R\'{e}nyi
entropies has zero measure (Bhattacharyya measure). In addition,
we show the instabilities can be easily emended by introducing a
coarse graining into an actual measurement. We also clear up any
doubts regarding the observability of R\'{e}nyi's entropy in
(multi--)fractal systems and in systems with absolutely continuous
probability density functions.
\\

\vspace{3mm}
\end{abstract}


\section{Introduction}

Thermodynamical or statistical concept of entropy, though deeply
rooted in physics, is rigorously defined only for equilibrium
systems or, at best, for adiabatically evolving systems. In fact,
the very existence of the entropy in thermodynamics is attributed
to Carath\'{e}odory's inaccessibility theorem~\cite{Car1} and the
statistical interpretation behind the thermodynamical entropy is
then usually provided via the ergodic hypothesis~\cite{Rb1,Khin2}.
It is, however, highly non--trivial matter to find a proper
conceptual ground for entropy of systems away from equilibrium,
non--ergodic systems or equilibrium systems with ``exotic"
non--Gibbsian statistics (multifractals, percolation, polymers or
protein folding provide examples). It is frequently said that
entropy is a measure of disorder, and while this needs many
qualifications and clarifications it is generally believed that
this does represent something essential about it. Information
theory might be then viewed as a pertinent mathematical framework
capable of quantifying the ``measure of disorder". It is undoubted
advantage of information theoretic approaches that whenever one
can measure (or control) information one can also measure (or
control) the associated entropy, as the latter is essentially an
average information about a system in question~\cite{Re1,Pj3}.

\vspace{3mm}

In recent years there have been many attempts to extend the
equilibrium concept of entropy to more generic situations applying
various generalizations of the information theory. Systems with
(multi--)fractal structure, long--range interactions and
long--time memories might serve as examples. Among multitude of
information entropies Shannon's entropy, R\'{e}nyi entropies and
Tsallis--Havrda--Charvat (THC) non--extensive entropies~\cite{THC}
have found utility in a wide range of physical problems. Shannon's
entropy is known to reproduce the usual Gibbsian thermodynamics
and is frequently used in such areas as astronomy, geophysics,
biology, medical diagnosis and economics (for the latest
developments in Shannon's entropy applications the interested
reader may consult Ref.~\cite{Max} and citations therein).
R\'{e}nyi entropies were conveniently applied, for instance, in
multiparticle hadronic systems~\cite{ABi1}, fractional diffusion
processes~\cite{CEs1} or in multifractal systems~\cite{TCH1}. THC
entropy was recently used in a study of systems with strong
long--range correlations and in systems with long--time
memories\cite{Ab1}.

\vspace{3mm}

Despite the information theoretic origin there has been raised
some doubt regarding the observability of R\'{e}nyi
entropies~\cite{Lesche}. Some authors went even as far as to claim
that instabilities in systems with large number of microstates
completely invalidate use of R\'{e}nyi entropies in all physical
problems~\cite{Abe2}. This is rather surprising since R\'{e}nyi's
entropy is routinely measured in numerous situation ranging from
 coding theory and
cryptography~\cite{Cac} (where it regulates the optimality of
coding), through chaotic dynamical systems~\cite{TCH1} (where it
determines the generalized dimensions for strange attractors) and
earthquake analysis~\cite{Harte1} (where it is used to evaluate
the distribution of earthquake epicenters and lacunarity) to
non--parametric mathematical statistics (where it prescribes the
price of constituent information). Besides, R\'{e}nyi entropies
directly provide measurable bounds in quantum--information
uncertainty relations~\cite{HM1}.

\vspace{3mm}

In the present paper we aim to revise Lesche's condition of
observability. We illustrate this in various contexts; systems
with finite number of microstates, systems with infinite (but
countable) number of microstates, systems with absolutely
continuous probability density functions (PDF's) and
multifractals. We show that it is not quite as simple to define
the ubiquitous concept of observability. We propose a less
restrictive observability condition and demonstrate that R\'{e}nyi
entropies are observable in this new framework. In what follows we
will give some considerations in favor of the above statement.

\vspace{3mm}

The paper is organized in the following way: In Section~\ref{les1}
we discuss Lesche's criterion of observability which frequently
forms a core argument against observability of R\'{e}nyi
entropies. We argue that the criterion is unnecessary restrictive
and, in fact, many standard physical phenomena which are observed
and measured in a real world do not obey Lesche's  condition. In
Section~\ref{Ren23} we present some essentials of R\'{e}nyi
entropies required in the main body of the paper. In
Section~\ref{finite1} we argue that for the finite number of
microstates R\'{e}nyi entropies easily conform with Lesche's
criterion, i.e., they are observable. In Section~\ref{infinite 1}
we extend our analysis to a countably infinite number of
microstates. Here appearance of instabilities may be observed. The
latter can be traced to a large sensitivity of R\'{e}nyi entropies
to (ultra)rare--event systems. We demonstrate that when the coarse
graining is included into realistic measurements, the
instabilities get ``diluted" and R\'{e}nyi entropies once again
obey Lesche's condition. In Section~\ref{infinite 3} we propose
more realistic criterion of observability where we allow for a
certain amount of instability points, provided the latter ones
have measure zero. To this extend we employ Bhattacharyya
statistical measure - i.e., natural measure on the space of
non--parametric statistics. We prove that the Bhattacharyya
measure of the above ``critical" distributions is in fact zero.
Finally, we analyze in Section~\ref{multifrac} systems with
continuous probability distributions and multifractal systems. We
find that the very nature of the absolute continuity of PDF's and
the multifractality prohibits {\em per se} an appearance of
instability points.


\section{Lesche's criterion of
observability}\label{les1}

In order to explain fully the apparent inconsistencies in the
recent claims concerning non--observability of R\'{e}nyi entropies
we feel it is necessary to briefly review the main points of
Lesche's observability criterion. While we hope to discuss all the
salient points, a full discussion can be obtained in
Lesche~\cite{Lesche}. Our discussion will be in terms of a scalar
quantity $G({\boldsymbol{x}})$. Following~\cite{Lesche}, a
necessary condition for $G({\boldsymbol{x}})$ with the
state\footnote{Here and throughout, the state space $X$ represents
the sample space of mathematical statistics, i.e., the space over
which the probability distributions operate. In simple situations
this coincides with the set of all possible outcomes in some
experiment. Generally, the elements of $X$ can represent
probability distributions themselves provided a suitable measure
is defined. This fact will be used in Section~IV.} variable
${\boldsymbol{x}} \in X\subset {\mathbb{R}}^n$ to be observable is
following: Let
\begin{eqnarray*}
||{\boldsymbol{x}} - {\boldsymbol{x}}'||_1 = \sum_k^n |x_k -
x'_k|\, ,
\end{eqnarray*}
\noi be the H\"{o}lder $l_1$--metric
on ${\mathbb{R}}^n $, then for $\forall \varepsilon >0$ there
exists (${\boldsymbol{x}}$--independent) $\delta_{\varepsilon}>0$
such that for any pair ${\boldsymbol{x}}$, ${\boldsymbol{x}}'$ one
has
\begin{eqnarray}
||{\boldsymbol{x}} - {\boldsymbol{x}}'||_1 \leq
\delta_{\varepsilon} \ \Rightarrow \ \frac{|G({\boldsymbol{x}}) -
G({\boldsymbol{x}}')|}{G_{max}} < \varepsilon\, . \label{lesh1}
\end{eqnarray}
\noi From a strict mathematical standpoint (\ref{lesh1}) is, in
fact, the definition of the uniform metric continuity of
$G({\boldsymbol{x}})$ on the state space $X$. Informally
Eq.(\ref{lesh1}) states that points from $X$ which are close in
sense of $||\ldots ||_1$ are mapped via $G$ to points which are
close in $|\ldots |$ metric. Lesche's criterion is thus nothing
but the condition of stability of $G({\boldsymbol{x}})$ under a
measurement. In fact, the continuity criterion ensures that a
small error in a state variable ${\boldsymbol{x}}$ will not bring
in repeated experiments violent fluctuations in measured data. The
{\em uniform} continuity in Eq.(\ref{lesh1}) is then a key
ingredient to secure that the size of the changes in
$G({\boldsymbol{x}})$ depends only on the size of the changes in
${\boldsymbol{x}}$ but not on ${\boldsymbol{x}}$ itself. This
condition excludes, for example, systems whose statistical
fluctuations in $G({\boldsymbol{x}})$ would change too
dramatically with a small change in the state variable
${\boldsymbol{x}}$.


\vspace{3mm}

When $G({\boldsymbol{x}})$ is bounded we can recast Lesche's
condition of observability into equivalent but more expedient
form, namely (inverse) Lipschitz continuity
condition~\cite{Rudin1}. In this case, a quantity $G:
X\subset{\mathbb{R}}^n \mapsto \mathbb{R}$ is observable in
Lesche's sense if and only if for every $\varepsilon >0$ there
exists (${\boldsymbol{x}}$--independent and finite)
$K_{\varepsilon}$ such that for any pair ${\boldsymbol{x}}$,
${\boldsymbol{x}}' \in X$ one has
\begin{equation}
|G({\boldsymbol{x}}) - G({\boldsymbol{x}}')| \leq K_{\varepsilon}
||{\boldsymbol{x}} - {\boldsymbol{x}}'||_1  + \varepsilon \, .
\label{lifs1}
\end{equation}
\noi We will practically employ the condition (\ref{lifs1}) in
Section~\ref{finite1}.

\vspace{3mm}

Criteria (\ref{lesh1}) and (\ref{lifs1}) get generalized in case
when $n \rightarrow \infty$. This should be expected as the
uniform continuity may not survive in the large $n$ limit. To
avoid such situations Lesche postulated that the mapping
\begin{eqnarray}
G: \bigcup_{n=1}^{\infty} X_n \mapsto {\mathbb{R}}\, ,
\end{eqnarray}
\noi with $X_n \subset {\mathbb{R}}^n$, taken as a function of
$n$, converges to an uniformly continuous function in an uniform
manner, i.e., for $\forall \varepsilon >0$ there exists
 $\delta_{\varepsilon} >0$
such that for $\forall$ ${\boldsymbol{x}}, {\boldsymbol{x}}' \in
{\mathbb{R}}^n$ and $\forall n \in {\mathbb{Z}}^+$
\begin{eqnarray}
||{\boldsymbol{x}} - {\boldsymbol{x}}'||_1 \leq
\delta_{\varepsilon} \ \Rightarrow \ \frac{|G({\boldsymbol{x}}) -
G({\boldsymbol{x}}')|}{G_{max}} < \varepsilon\, . \label{lifs2}
\end{eqnarray}
\noi The uniform convergence is then reflected in the fact that
$\delta_{\varepsilon}$ is both ${\boldsymbol{x}}$ and $n$
independent.

\vspace{3mm}

Let us add a couple of remarks concerning the aforementioned
observability conditions. Lesche's condition, as illustrated
above, is based on the notion of measurability. This is however
not the only possible way how to define observability. It is well
known that various alternative concepts exist in literature. For
instance, one may use the approach based on
distinguishability~\cite{Ber1} or detectability~\cite{Sontag}. In
fact, the condition based on measurability, and namely the
condition of uniform continuity, might be often to tight. Indeed,
there are clearly many quantities which are not uniformly
continuous in their state variables (e.g., they are discontinuous
in finite number of points in the state space) and which are,
nevertheless, perfectly detectable and well defined away from the
singularity domain. Note, for instance, that although pressure and
latent heat in 1st order phase transitions are discontinuous in
temperature, and similarly susceptibility in 2nd order phase
transitions is nonanalytic in temperature, there is still no
reason to dismiss pressure, latent heat and susceptibility as
observables. Discontinuous or nonanalytic {\em state} functions
are not exclusive to phase transitions only. Actually, such a type
of behavior is common to many different situations - formation of
shocks in nonlinear wave propagation, mechanical systems involving
small masses and large damping, electric--circuit systems with
large resistance and small inductance, catastrophe and bifurcation
theories, to name a few.

\section{R\'{e}nyi entropies}\label{Ren23}

R\'{e}nyi entropies constitute a one--parametric family of
information entropies labelled by R\'{e}nyi's parameter $\alpha
\in {\mathbb{R}}^+$ and fulfill the additivity with respect to the
composition of statistically independent systems. The special case
with $\alpha = 1$ corresponds to ordinary Shannon's entropy. It
might be shown that R\'{e}nyi entropies belong to the class of
mixing homomorphic functions~\cite{Lesche} and that they are
analytic for $\alpha$'s which lie in  $I \cup IV$ quadrants of the
complex plane~\cite{PJ1}. In order to address the observability
issue it is important to distinguish three situations.

\subsection{Discrete probability distribution case }

Let ${\mathcal{X}} = \{ x_1, \ldots, x_n \}$ be a random variable
admitting $n$ different events (be it outcomes of some experiment
or microstates of a given macrosystem), and let ${\mathcal{P}} =
\{ p_1, \ldots, p_n \} $ be the corresponding probability
distribution. Information theory then ensures that the most
general information measures (i.e., entropy) compatible with the
additivity of independent events are those of
R\'{e}nyi~\cite{Re1}:
\begin{equation}
{\cal{I}}_\alpha({\cal{P}}) = \frac{1}{(1-\alpha)}\, \mbox{log}_2
\left( \sum_{k=1}^{n}p^{\alpha}_k\right) \, .  \label{ren11}
\end{equation}
\noi Form (\ref{ren11}) is valid even in the limiting case when
$n\rightarrow \infty$. If, however, $n$ is finite then R\'{e}nyi
entropies are bounded both from below and from above:
$\log_2(p_k)_{\mbox{\scriptsize{max}}}  \leq {\cal{I}}_{\alpha}
\leq \log_2 n.$ In addition, R\'{e}nyi entropies are monotonically
decreasing functions in $\alpha$, so namely ${\cal{I}}_{\alpha_1}
< {\cal{I}}_{\alpha_2}$  if and only if $\alpha_1
> \alpha_2$. One can reconstruct the entire underlying
probability distribution knowing all R\'{e}nyi distributions via
Widder--Stiltjes inverse formula~\cite{PJ1}. In the latter case
the leading order contribution comes from
${\cal{I}}_1({\cal{P}})$, i.e., from Shannon's entropy. Typical
playground of (\ref{ren11}) is in a coding theory~\cite{Cam1},
cryptography~\cite{Cac} and in theory of statistical
inference~\cite{Re1}. The parameter $\alpha$ might be then related
with the price of constituent information. It should be admitted
that in discrete cases the conceptual connection of
${\cal{I}}_\alpha({\cal{P}})$ with actual physical problems is
still an open issue. The interested reader can find some further
practical applications of discrete R\'{e}nyi entropies, for
instance, in~\cite{PJ1,Sa1}

\subsection{Continuous probability distribution case \label{contin1}}

Let $M$ be a support on which is defined a continuous PDF
${\mathcal{F}}({\boldsymbol{x}})$. We will assume that the support
(or outcome space) can be generally a fractal set. By covering the
support with the mesh $M^{(l)}$ of $d$--dimensional (disjoint)
cubes $M^{(l)}_k$ $(k=1, \ldots, n)$ of size $l^d$ we may define
the integrated probability in $k$--th cube as
\begin{equation}
p_{nk} = {\mathcal{F}}({\boldsymbol{x}}_i) l^d\, , \;\;\;
{\boldsymbol{x}}_i \in M_{k}^{(l)}\, .
\end{equation}
\noi The latter specifies the mesh probability distribution
${\mathcal{P}}_n =\{ p_{n1}, \ldots, p_{nn} \}$. Infinite
precision of measurements (i.e., with $l \rightarrow 0$) often
brings infinite information. In fact, it is more sensible to
consider the relative information entropy rather than absolute one
as the most ``junk" information comes from the uniform
distribution ${\mathcal{E}}_n$. It was shown in~\cite{Re1,PJ1}
that in the $n \rightarrow \infty$ (i.e., $l \rightarrow 0$) limit
it is possible to define finite information measure compatible
with information theory axioms. This {\em renormalized}
R\'{e}nyi's entropy - negentropy (or information gain), reads
\begin{eqnarray}
\tilde{{\cal{I}}}_{\alpha}({\cal{F}}) &\equiv& \lim_{n \rightarrow
\infty} ({\cal{I}}_{\alpha}({\cal{P}}_n) -
{\cal{I}}_{\alpha}({\cal{E}}_n) ) \ = \ \frac{1}{(1-\alpha)} \,
\log_2 \left(\frac{\int_M d \mu \,
{\cal{F}}^{\alpha}({\boldsymbol{x}})}{\int_M d \mu \,
1/V^{\alpha}} \right) \, . \label{ren55}
\end{eqnarray}
\noi Here $V$ is the corresponding volume. Eq.(\ref{ren55}) can be
viewed as a generalization of the Kullback--Leibler relative
entropy~\cite{KL1}. It is possible to introduce a simpler
alternative prescription as
\begin{eqnarray}
{\cal{I}}_{\alpha}({\cal{F}}) &\equiv& \lim_{n \rightarrow \infty}
({\cal{I}}_{\alpha}({\cal{P}}_n)
-{\cal{I}}_{\alpha}({\cal{E}}_n)|_{V=1} )\ = \ \lim_{n \rightarrow
\infty} ({\cal{I}}_{\alpha}({\cal{P}}_n) + D \log_2 l) \nonumber
\\ &=& \frac{1}{(1-\alpha)} \log_2 \left( \int_M d \mu \,
{\cal{F}}^{\alpha}({\boldsymbol{x}}) \right)\, . \label{ren6}
\end{eqnarray}
\noi In both previous cases the measure $\mu $ is the Hausdorff
measure~\cite{Fed1}:
\begin{eqnarray*}
\mu(d;l) = \sum_{k{\rm th~box}} {l^d} \ \stackrel{l \rightarrow
0}{\longrightarrow} \ \left\{
\begin{array}{ll}
0 & \mbox{if $d < D$}\\
\infty & \mbox{if $d > D$} \, ,
\end{array} \right.
\end{eqnarray*}
\noi with $D$ being the Hausdorff dimension of the support.
R\'{e}nyi entropies (\ref{ren55}) and (\ref{ren6}) are defined if
and only if the corresponding integral $\int_M d \mu
\,{\cal{F}}^{\alpha}({\boldsymbol{x}})$ exists. Eqs.(\ref{ren55})
and (\ref{ren6}) indicate that asymptotic expansion for
${\cal{I}}_{\alpha}({\cal{P}}_n)$ has the form:
\begin{eqnarray}
{\cal{I}}_{\alpha}({\cal{P}}_n) \ = \ - D\log_2 l +
{\mathcal{I}}_{\alpha}({\mathcal{F}}) + o(1)\ = \ - D\log_2 l +
\tilde{{\cal{I}}}_{\alpha}({\cal{F}}) + \log_2 V_n + o(1)\, .
\label{as1}
\end{eqnarray}
\noi Here $V_n$ is the pre--fractal volume and the symbol $o(1)$
is the residual error which tends to $0$ for $l \rightarrow 0$. In
contrast to the discrete case, R\'{e}nyi entropies
${\mathcal{I}}_{\alpha}({\mathcal{F}})$ are not positive here.

\vspace{3mm}

Information measures $\tilde{\mathcal{I}}_{\alpha}({\mathcal{F}})$
and ${\mathcal{I}}_{\alpha}({\mathcal{F}})$ have been so far
mostly applied in theory of statistical inference~\cite{TA1} and
in chaotic dynamical systems~\cite{TCH1}. Let us note finally that
one may view the discrete distributions as a special case of the
continuous PDF's, provided the outcome space (or sample space) is
discrete. In such a situation the Hausdorff dimension $D$ is zero
and Eq.(\ref{ren6}) reduces directly to Eq.(\ref{ren11}).

\subsection{Multifractal systems}

Multifractals can be viewed as statistical systems where both
cells in the covering mesh and integrated probabilities scale as
some power of $l$. Grouping all the integrated probabilities
according to their scaling exponents (Lipshitz--H\"{o}lder
exponents), say $a$, we effectively divide the support into the
ensemble of intertwined unifractals each with its own fractal
dimension $f(a)$. Exponents $f(a)$ are called singularity
spectrum. In multifractal analysis it is customary to introduce
yet another pair of scaling exponents, namely the correlation
exponent $\tau(\alpha)$ which prescribes scaling of the partition
function and ``inverse temperature" $\alpha$. These two
descriptions are related via Legendre transformation
\begin{equation}
\tau(\alpha) = \min_{a} (\alpha a - f(a))\, .
\end{equation}
\noi As in the case of continuous PDF's the renormalization of
R\'{e}nyi entropies is required to extract relevant finite
information - negentropy. It is possible to show that the
following renormalized R\'{e}nyi's entropy complies with the
axiomatics of the information theory\cite{PJ1}:
\begin{eqnarray}
{\mathcal{I}}_{\alpha}(\mu_{{\mathcal{P}}}) &\equiv& \lim_{l
\rightarrow 0} \ \left({\mathcal{I}}_{\alpha}({\mathcal{P}}_n) -
{\mathcal{I}}_{\alpha}({\mathcal{E}}_n) |_{V=1} \right)\nonumber \\
&=& \lim_{l \rightarrow 0} \
\left({\mathcal{I}}_{\alpha}({\mathcal{P}}_n) +
\frac{\tau(\alpha)}{(\alpha -1)} \log_2 l \right) \nonumber \\
&=& \frac{1}{(1-\alpha)} \ \log_2\left( \int_a
d\mu_{{\mathcal{P}}}^{(\alpha)}(a)\right)\, . \label{mes6}
\end{eqnarray}
\noi Here the multifractal measure is defined as~\cite{Fed1}
\begin{eqnarray*}
\mu_{{\mathcal{P}}}^{(\alpha)}(d;l) = \sum_{k{\rm th~box}}
\frac{p^{\alpha}_{nk}}{l^d} \ \stackrel{l \rightarrow
0}{\longrightarrow} \ \left\{
\begin{array}{ll}
0 & \mbox{if $d < \tau(\alpha)$}\\
\infty & \mbox{if $d > \tau(\alpha)$}\, .
\end{array}
\right.
\end{eqnarray*}
\noi R\'{e}nyi entropies
${\mathcal{I}}_{\alpha}(\mu_{{\mathcal{P}}})$ are defined if and
only if the corresponding integrals $\int_a
d\mu_{{\mathcal{P}}}^{(\alpha)}(a)$ exist. Eq.(\ref{mes6}) implies
the following asymptotic expansion for
${\mathcal{I}}_{\alpha}({\mathcal{P}}_n)$
\begin{equation}
{\mathcal{I}}_{\alpha}({\mathcal{P}}_n) = - D(\alpha) \log_2 l +
{\mathcal{I}}_{\alpha}(\mu_{{\mathcal{P}}}) + o(1)\, . \label{as2}
\end{equation}
\noi Here
\begin{equation}
D(\alpha) \equiv  \frac{\tau(\alpha)}{(\alpha -1 )} = \lim_{l
\rightarrow 0} \frac{{\mathcal{I}}_{\alpha}({\mathcal{P}}_n)}{
\log_2 (1/l)}\, ,
\end{equation}
\noi is the, so called, generalized dimension~\cite{Fed1}\!. Note
also that for systems of Subsection~\ref{contin1}  $D(\alpha)$ is
$\alpha$ independent.

\vspace{3mm}

Let us stress that R\'{e}nyi's entropy of multifractal systems is
more convenient tool than the ordinary Shannon's entropy. It is
possible to show that one can obtain Shannon's entropy for any
unifractal by merely changing the R\'{e}nyi parameter. In fact,
R\'{e}nyi's parameter coincides in this case with the singularity
spectrum~\cite{PJ1}.

\section{Observability of R\'{e}nyi entropies: Discrete
probability distribution}\label{dis1}

\subsection{Finite case} \label{finite1}

It is quite simple to see that for systems with a finite number of
outcomes (e.g., systems with finite number of microstates)
Lesche's criterion of observability is fulfilled. The proof goes
as follows\footnote{For simplicity's sake we use in this
subsection a natural logarithm instead of $\log_2$.}. We first use
the inequality $\lg x \ \leq \ x-1$ and assume that $\sum_k
p_k^{\alpha} \geq \sum_k q_k^{\alpha}$, then
\begin{eqnarray*}
\left| {\mathcal{I}}_{\alpha}({\mathcal{P}})-
{\mathcal{I}}_{\alpha}({\mathcal{Q}})  \right| \ \leq \
\frac{1}{|1-\alpha|} \left( \frac{\sum_{i=1}^n
p_i^{\alpha}}{\sum_{i=1}^n q_i^{\alpha}} - 1 \right)  =
\frac{1}{|1-\alpha| \ \sum_{i=1}^n q_k^{\alpha}} \  \sum_{i=1}^n
(p_i^{\alpha} - q_i^{\alpha})  \, .
\end{eqnarray*}
\noi This might be written in the invariant form as
\begin{eqnarray}
\left| {\mathcal{I}}_{\alpha}({\mathcal{P}})-
{\mathcal{I}}_{\alpha}({\mathcal{Q}})  \right| & \leq &
\frac{1}{|1-\alpha| \ c(\alpha,{\mathcal{P}},{\mathcal{Q}} )} \
\left| \sum_{i=1}^n  (p_i^{\alpha} - q_i^{\alpha})  \right|
\nonumber \\  &\leq & \frac{1}{|1-\alpha| \ d(\alpha,n )} \ \left|
\sum_{i=1}^n (p_i^{\alpha} - q_i^{\alpha})  \right|\, .
\label{discrete1}
\end{eqnarray}
\noi Here $c(\alpha,{\mathcal{P}},{\mathcal{Q}} ) =
\min\left(\sum_i p_i^{\alpha}, \sum_i q_i^{\alpha} \right)$ and
\begin{eqnarray*}
&&  d(\alpha, n) = \left\{ \begin{array}{ll}
                          1 & \mbox{if $0 < \alpha \leq1$}\\
                          n^{1-\alpha} & \mbox{if $\alpha\geq1$}\,
                          .
                          \end{array}\right.
\end{eqnarray*}
\noi To find the efficient estimate for $\left| \sum_k
(p_k^{\alpha} - q_k^{\alpha}) \right|$ in terms of $||
{\mathcal{P}} - {\mathcal{Q}}||_1$ we utilize the following trick:
Let us define the function
\begin{eqnarray}
{\mathcal{A}}(s,{\mathcal{P}}) = \sum_{k=1}^{n}(p_k -
f(s))\theta(p_k - f(s))\, .
\end{eqnarray}
\noi Here $\theta(\ldots)$ is the Heaviside step function and $f:
[a,b] \mapsto [0,1]$ is some invertible function. Both $f(s)$, $a$
and $b$ will be chosen at the latter stage so as to facilitate our
computations. Note also that
\begin{eqnarray}
\max\left\{0; (1 - nf(s))\right\}
\leq{\mathcal{A}}(s,{\mathcal{P}}) \leq 1\, . \label{ineq0}
\end{eqnarray}
\noi Important property of ${\mathcal{A}}(s,{\mathcal{P}})$ is the
following straightforward inequality
\begin{eqnarray}
|{\mathcal{A}}(s,{\mathcal{P}}) - {\mathcal{A}}(s,{\mathcal{Q}})|
&\leq& \sum_{k=1}^n |(p_k - f(s))\theta(p_k - f(s)) - (q_k -
f(s))\theta(q_k - f(s)) |\nonumber \\
&\leq& \sum_{k=1}^n |p_k - q_k| = ||{\mathcal{P}} -
{\mathcal{Q}}||_1\, , \label{ineq1}
\end{eqnarray}
\noi which is valid for any $s\in [a,b]$. Note further that
\begin{eqnarray}
\int_a^{b} {\mathcal{A}}(s,{\mathcal{P}}) \ ds &=& \sum_{k=1}^n
\int_{f(a)}^{f(b)} (p_k - x)\theta(p_k -
x) \left( f^{-1}(x) \right)' dx \nonumber \\
&=& \sum_{k=1}^n \left\{ \theta(p_k - f(a))\left( (f(a) - p_k)a +
\int_{f(a)}^{p_k} f^{-1}(x) \ dx  \right) \right. \nonumber
\\ &&+ \ \left. \theta(p_k - f(b))\left((p_k - f(b))b + \int_{p_k}^{f(b)} f^{-1}(x)
\ dx \right) \right\}\, . \label{petr}
\end{eqnarray}
\noi Here we have used the fact that $p_k$'s must lie somewhere
between $f(a)$ and $f(b)$. If we now chose $f(x) =
(x/\alpha)^{1/(\alpha -1)}$ with
\begin{eqnarray*}
&&  a = \left\{ \begin{array}{ll}
                          \infty & \mbox{if $0 < \alpha <1$}\\
                          0 & \mbox{if $\alpha > 1$}
                          \end{array}\right. \!\! , \; \; \mbox{and} \;\; b = \alpha \, ,
\end{eqnarray*}
\noi (so $f(a) = 0, f(b) = 1$) we obtain
\begin{eqnarray}
\left|\int_a^{b} \left({\mathcal{A}}(s,{\mathcal{P}}) -
{\mathcal{A}}(s,{\mathcal{Q}})\right) ds \right| =
\left|\sum_{k=1}^n(p_k^{\alpha} - q_k^{\alpha} ) \right|\, .
\label{ineq2}
\end{eqnarray}

\noi Applying (\ref{ineq1}) and (\ref{ineq2}) we may write for
$\alpha > 1$
\begin{eqnarray}
\left|\sum_{k=1}^n(p_k^{\alpha} - q_k^{\alpha} ) \right| &\leq& \
\left\{\int_{0}^{c} n(s/\alpha)^{1/(\alpha -1)}\ ds +
\int_c^{\alpha}\left| {\mathcal{A}}(s,{\mathcal{P}}) -
{\mathcal{A}}(s,{\mathcal{Q}}) \right| \ ds \
 \right\}
\nonumber \\
&\ \leq& \ n(\alpha -1) (c/\alpha)^{\alpha/\alpha -1} + (\alpha -
c)||{\mathcal{P}} - {\mathcal{Q}}||_1 \, .
\end{eqnarray}

\noi So if we take $c = \alpha (\varepsilon/n^\alpha)^{(\alpha
-1)/\alpha}$ (this assures that ${\mathcal{A}}(s,{\mathcal{P}})
\geq (1-nf(s))
> 0$ for $s \in (0,c]$) then
\begin{eqnarray}
\left| {\mathcal{I}}_{\alpha}({\mathcal{P}}) -
{\mathcal{I}}_{\alpha}({\mathcal{Q}}) \right| \ \leq  \
K_{\varepsilon}^{(1)}\ ||{\mathcal{P}} - {\mathcal{Q}}||_1 +
\varepsilon \, , \label{lifs3}
\end{eqnarray}

\noi with $K_{\varepsilon}^{(1)} = \frac{\alpha}{(\alpha
-1)}\left( (n^{\alpha}/\varepsilon)^{(\alpha -1)/\alpha} -
1\right)\varepsilon^{(\alpha -1)/\alpha}$.

\vspace{3mm}

In case when $0< \alpha < 1$ we may utilize (\ref{ineq0}),
(\ref{ineq1}) and (\ref{ineq2}) to obtain
\begin{eqnarray}
\left|\sum_{k=1}^n(p_k^{\alpha} - q_k^{\alpha} ) \right| &\leq& \
\left\{\int_{\alpha}^{\tilde{c}} \left|
{\mathcal{A}}(s,{\mathcal{P}}) - {\mathcal{A}}(s,{\mathcal{Q}})
\right| ds + \int_{\tilde{c}}^{\infty}n(s/\alpha)^{1/(\alpha - 1)}
\ ds \
 \right\}
\nonumber \\
&\leq& \ (\tilde{c}-\alpha)  || {\mathcal{P}} - {\mathcal{Q}}||_1
+ n(1-\alpha )  (\tilde{c}/\alpha)^{\alpha/(\alpha -1)}\, \, .
\end{eqnarray}

\noi By setting $\tilde{c} = \alpha (\varepsilon/n)^{(\alpha
-1)/\alpha}$ (this assures that ${\mathcal{A}}(s,{\mathcal{P}})
\geq (1-nf(s))
> 0$ for $s \in [\tilde{c},\infty)$) we have
\begin{eqnarray}
\left| {\mathcal{I}}_{\alpha}({\mathcal{P}})-
{\mathcal{I}}_{\alpha}({\mathcal{Q}})  \right|
&\leq & K_{\varepsilon}^{(2)}\ || {\mathcal{P}} -
{\mathcal{Q}}||_1 + \varepsilon  \, , \label{lifs4}
\end{eqnarray}

\noi with $K_\varepsilon^{(2)} = \frac{\alpha}{(1-\alpha)} \left(
(\varepsilon/n)^{(\alpha -1)/\alpha} - 1  \right)$. Note
particularly that $\lim_{\alpha \rightarrow 1_+}
K_{\varepsilon}^{(1)} = \ln(n/\varepsilon)$ and
$\lim_{\alpha\rightarrow 1_-} K_\varepsilon^{(2)} =
\ln(n/\varepsilon)$. This indicates that the Lipschitz conditions
(\ref{lifs3}) and (\ref{lifs4}) can be analytically continued to
$\alpha =1$. This reconfirms the well known result that Shannon's
entropy is Lipschitz.

\vspace{3mm}

Finally note that Eqs.(\ref{lifs3}) and (\ref{lifs4}) represent
the Lesche criterion (\ref{lifs1}). Hence, in cases when the state
space corresponds to the space of all possible probability
distributions
assigned to a definite (finite) number of outcomes (microstates)
R\'{e}nyi entropies are measurable in Lesche's sense.
%

\subsection{Infinite limit case} \label{infinite 1}

As it was already mentioned in Section~\ref{les1}, the situation
starts to be more delicate in the large $n$ limit. This is because
for the sake of uniform metric continuity at any $n$ one might
require that also the limiting case should obey the uniform
continuity. To tackle
statistical systems with a countable infinity of
microstates\footnote{Such systems often appear in various physical
situations. (Countable) Markov chains, Fermi-Pasta-Ulam lattice
models or symbolic dynamical models being examples.} we will
illustrate first that by introducing a coarse graining into a
realistic measurement, alleged Lesche's counterexamples do not
apply.

\vspace{3mm}

In his paper~\cite{Lesche} Lesche proposed the following examples
to demonstrate the non--observability of R\'{e}nyi entropies. In
$\alpha > 1$ he picked up two distributions, namely ($i = 1,
\ldots, n$)
\begin{eqnarray}
&&{\mathcal{P}} = \left\{ p_i = \frac{1}{n-1} \left( 1-
\delta_{1i}\right) \right\}\, , \nonumber \\ && {\mathcal{P}}' =
\left\{ p_i^{\ ,} = \frac{\delta}{2} \ \delta_{1i} + \left(1 -
\frac{\delta}{2}\right)  \left(\frac{1-\delta_{1i}}{n-1}
\right)\right\}\, , \nonumber \\ && || {\mathcal{P}} -
{\mathcal{P}}' ||_1 = \delta \, . \label{lesh11}
\end{eqnarray}
\noi Lesche then went on to show that these two distributions do
not fulfill the uniform continuity in the large $n$ limiting case.
Let us now show that the coarse graining (which is naturally
present in any realistic measurement) will restore the uniform
continuity for the large $n$ limit case.

\vspace{3mm}

We will assume, for the simplicity's sake, that the discrete
probability distributions (\ref{lesh11}) are living on the unit
lattice with equidistantly distributed lattice (i.e., support)
points. In the spirit of Lesche's paper we assume that the true
probability distribution on the interval $[0,1]$ is obtained in
$n\rightarrow\infty$ limit (i.e., when the lattice spacing tends
to zero). As usually, we will keep $n\gg 1$ finite during
calculations and set to infinity only at the very last stage.
Because every actual measurements have a certain resolution
capacity we will further assume that a realistic measurement can
sample the unit interval through window of width $1/k$ ($k \ll n$)
(so $k$ windows will cover the support space). In this case one
can know only integrated probabilities, hence ${\mathcal{P}}
\rightarrow {\mathcal{P}}_{(k)}$ and ${\mathcal{P}}' \rightarrow
{\mathcal{P}}'_{(k)}$. As in every window there is $n/k$
underlying $p_i$'s we have ($i = 1, \ldots, k$)
\begin{eqnarray}
&& {\mathcal{P}}_{(k)} = \left\{p_i^{(k)} = \frac{1}{n-1}\left(
\frac{n}{k} -\delta_{1i} \right) \right\} \, , \nonumber \\
&& {\mathcal{P}}'_{(k)} = \left\{ p_i^{\ ,(k)} = \frac{\delta}{2}\
\delta_{1i} + \frac{\left( 1- \delta/2   \right)}{n-1}\
\left(\frac{n }{k} - \delta_{1i} \right) \right\}\, , \nonumber
\\
&& ||{\mathcal{P}}_{(k)} - {\mathcal{P}}'_{(k)}||_1 = \delta \, .
\label{leshe4}
\end{eqnarray}
\noi Using the fact that ${\mathcal{I}}_{\alpha \ {\rm max}} =
\log_2 k$ we have
\begin{eqnarray}
\frac{\left|{\mathcal{I}}_{\alpha}({\mathcal{P}}_{(k)}) -
{\mathcal{I}}_{\alpha}({\mathcal{P}}'_{(k)})
\right|}{{\mathcal{I}}_{\alpha \ {\rm max}}}\!\! &=&\!\! \left|
\frac{1}{(1 -\alpha)} \ \log_2 \left[ \frac{\left( \frac{1}{n-1}
\right)^{\alpha}\left(\frac{n}{k} -1\right)^{\alpha}  +
(k-1)\left( \frac{1}{n-1} \right)^{\alpha} \left(\frac{n}{k}
\right)^{\alpha}}{\left(\frac{\delta}{2} + \frac{(1 - \ \delta/2
)}{n-1} \left( \frac{n}{k} -1 \right) \right)^{\alpha}  + (k-1)
\left( \frac{1 - \ \delta/2}{n-1}  \right)^{\alpha} \left(
\frac{n}{k}\right)^{\alpha} } \right] \right| \nonumber \\
&& \times \ (\log_2 k)^{-1} \nonumber \\
&~&  \nonumber \\
& \stackrel{n \rightarrow \infty}{\longrightarrow} & \left|
\frac{1}{( 1- \alpha )} \ \log_2 \left[ \frac{\left( \frac{1}{k}
\right)^{\alpha} + (k-1) \left( \frac{1}{k} \right)^{\alpha}
}{\left( \frac{\delta}{2} + \frac{(1-  \ \delta/2)}{k}
\right)^{\alpha}  + (k-1) \left( \frac{1- \ \delta/2}{k}
\right)^{\alpha}} \right]/ \log_2 k \right| \nonumber \\ &~& \nonumber \\
&=& \!\! \left|\frac{1}{1- \alpha } - \frac{1}{1-
\alpha}\log_2\left[ \left(1 + \frac{\delta}{2}(k-1)
\right)^{\alpha} + (k-1)\left( 1 - \frac{\delta}{2}
\right)^{\alpha}\ \right]/\log_2 k
\right|\nonumber \\ &~& \nonumber \\
&=& \left(\frac{\delta}{2} \right)^2 \frac{\alpha}{2}\
\frac{(k-1)}{\ln k} + {\mathcal{O}}(\delta^3)\, .
\end{eqnarray}
\noi It is now simple to see that Lesche's condition is easily
fulfilled, as for arbitrarily small $\varepsilon$ there exist
$\delta_{\varepsilon}$, namely
\begin{equation}
\delta_{\varepsilon} \ \leq  \ 2\sqrt{\frac{\varepsilon}{k-1}\ \ln
(k)^{2/\alpha}}\, , \label{leshe2}
\end{equation}
\noi for which the metric proximity $||{\mathcal{P}}_{(k)} -
{\mathcal{P}}'_{(k)}||_1 \leq \delta_{\varepsilon}$ implies the
proximity of outcomes, i.e.,
$\left|{\mathcal{I}}_{\alpha}({\mathcal{P}}_{(k)}) -
{\mathcal{I}}_{\alpha}({\mathcal{P}}'_{(k)}) \right|/\log_2 k \leq
\varepsilon$. This result is clearly independent of $n$ because
whenever $n$ is finite the outcome of the previous section applies
and for $n \rightarrow \infty$ the validity has been just proven.

\vspace{3mm}

We proceed analogously for $\alpha <1$. In this case Lesche's
counterexamples were provided by two distributions ($i = 1,
\ldots, n$)
\begin{eqnarray}
&& {\mathcal{P}} = \left\{ p_i = \delta_{1i} \right\}\, ,
\nonumber \\&& {\mathcal{P}}' = \left\{ p_i' = \left(1-
\frac{\delta}{2}\right)\ \delta_{1i}  + \frac{1}{n-1} \
\frac{\delta}{2}\ \left(1 - \delta_{1i} \right) \right\}\,
,\nonumber \\ && ||{\mathcal{P}} - {\mathcal{P}}'||_1 = \delta\, .
\label{example2}
\end{eqnarray}
\noi As before we can obtain integrated probability distributions
which read ($i = 1, \ldots, k$)
\begin{eqnarray}
&&{\mathcal{P}}_{(k)} = \left\{ p_i^{(k)} = \delta_{1i} \right\}\,
,
\nonumber \\
&&{\mathcal{P}}'_{(k)} = \left\{ {p'}_i^{(k)} = \left( 1-
\frac{\delta}{2}\right)\delta_{1i} + \frac{1}{n-1} \
\frac{\delta}{2}  \ \left( \frac{n}{k} - \delta_{1i}\right)
\right\}\, , \nonumber
\\&&||{\mathcal{P}}_{(k)} - {\mathcal{P}}'_{(k)}||_1 = \delta\, ,
\end{eqnarray}
\noi and so
\begin{eqnarray}
\frac{\left|{\mathcal{I}}_{\alpha}({\mathcal{P}}_{(k)}) -
{\mathcal{I}}_{\alpha}({\mathcal{P}}'_{(k)})
\right|}{{\mathcal{I}}_{\alpha \ {\rm max}}} &=& \left|
\frac{1}{(1-\alpha)} \ \log_2 \left[ \left(1 - \frac{\delta}{2}\
\frac{n(k-1)}{k(n-1)} \right)^{\alpha} + (k-1)\ \left(
\frac{\delta}{2}\  \frac{n}{k(n-1)} \right)^{\alpha} \right]
\right|\nonumber
\\
&& \times \ (\log_2 k)^{-1}\nonumber \\
&~& \nonumber \\
& \stackrel{n \rightarrow \infty}{\longrightarrow} &
\frac{1}{(1-\alpha)}\ \left| \log_2 \left[\left( 1-
\frac{\delta}{2}\ \frac{k-1}{k} \right)^{\alpha} + (k-1)
\left(\frac{\delta}{2 k}  \right)^{\alpha}
 \right]\right|/\log_2 k\nonumber
\\
&~& \nonumber \\
&\leq& \left( \frac{\delta}{2k} \right)^2 \frac{\alpha}{2} \
\frac{(k-1)^2}{\ln k}  + {\mathcal{O}}(\delta^3)\, .
\end{eqnarray}
\noi Here the inequality
\begin{eqnarray*}
x^\alpha - \alpha x \geq 0, \,\,\, \mbox{for} \,\,\, x \in [0,1],
\ \alpha \in [0,1]\, ,
\end{eqnarray*}
\noi was used on the last line. Consequently we again see that for
sufficiently small $\varepsilon$ there exist
$\delta_{\varepsilon}$, namely
\begin{equation}
\delta_{\varepsilon} \ \leq  \ \frac{2k}{(k-1)} \sqrt{\varepsilon
\ \ln (k)^{2/\alpha}}\, , \label{leshe3}
\end{equation}
\noi which satisfies Lesche's condition. Note, that from
(\ref{leshe2}) and (\ref{leshe3}) follows that our argument
naturally includes also the case $\alpha = 1$ (i.e., Shannon's
entropy) as in all steps leading to (\ref{leshe2}) and
(\ref{leshe3}) we have well defined limits $\alpha \rightarrow
1_+$ and $\alpha \rightarrow 1_-$, respectively.

\subsection{Region of instability} \label{infinite 3}

In the previous section we have found that Lesche's
counterexamples can be bypassed by introducing a coarsened
resolution into a measurement process. Let us now show that even
when the coarsening is not employed the Lesche instability points
have zero measure in the space of all discrete infinite
distributions - Bhattacharyya's measure~\cite{Bha1} - and hence
they  do not affect a measurement in most practical situations.

\vspace{3mm}

The key observation is that Lesche's counterexamples single out a
very narrow class of probability distributions. In particular,
they imply that when $\alpha
> 1$, only distributions with a high peak probabilities
create problems. Similarly, in cases where $\alpha < 1$ only
distributions with an infinite number of microstates having a
negligible overall probability exhibit a critical type of
behavior. We now demonstrate that the above probability
distributions have a very small relevance in actual measurement.
For this purpose we remind the reader the concept of Bhattacharyya
measure~\cite{Bha1}.

\vspace{3mm}

Suppose that ${\mathcal{X}}$ is a discrete random variable with
$n$ different values, ${\mathbb{P}}_{\!n}$ is the probability
space affiliated with ${\mathcal{X}}$ and ${\mathcal{P}}=\{p_1,
\ldots, p_n \}$ is a sample probability distribution from
${\mathbb{P}}_{\!n}$. Because ${\mathcal{P}}$ is non--negative and
summable to unity, it follows that the square--root likelihood
$\xi_i = \sqrt{p_i}$ exists for all $i = 1, \ldots, n$, and it
satisfies the normalization condition
\begin{eqnarray}
\sum_{i=1}^{n}(\xi_i)^2 =1\, .
\end{eqnarray}
We see that ${\boldsymbol{\xi}}$ can be regarded as a unit vector
in the Hilbert space ${\mathcal{H}} = {\mathbb{R}}^n$. Now, let
${\mathcal{P}}^{(1)}$ and ${\mathcal{P}}^{(2)}$ denote a pair of
probability distributions and ${\boldsymbol{\xi}}^{(1)}$ and
${\boldsymbol{\xi}}^{(2)}$ the corresponding elements in Hilbert
space. Then the inner product
\begin{eqnarray}
\cos \phi = \sum_{i=1}^n \xi^{(1)}_{i} \xi^{(2)}_{i} = 1-
\frac{1}{2} \sum_{i=1}^n \left(\xi^{(1)}_{i} - \xi^{(2)}_{i}
\right)^2\, ,
\end{eqnarray}
defines the angle $\phi$ that can be interpreted as a distance
between two probability distributions. More precisely, if
${\mathcal{S}}^{n-1}$ is the unit sphere in the $n$--dimensional
Hilbert space, then $\phi$ is the spherical (or geodesic) distance
between the points on ${\mathcal{S}}^{n-1}$ determined by
${\boldsymbol{\xi}}^{(1)}$ and ${\boldsymbol{\xi}}^{(2)}$.
Clearly, the maximal possible distance, corresponding to
orthogonal distributions, is given by $\phi = \pi/2$. This follows
from the fact that ${\boldsymbol{\xi}}^{(1)}$ and
${\boldsymbol{\xi}}^{(2)}$ are non--negative, and hence they are
located only on the positive orthant of ${\mathcal{S}}^{n-1}$.
Spherical geometry on ${\mathcal{S}}^{n-1}$ then naturally induces
the measure - Bhattacharyya measure.  The corresponding geodesic
distance $\phi$ is the, so called, Bhattacharyya distance. We
remark that the surface ``area" of the orthant
$({\mathcal{S}}^{n-1})^+$, i.e., the volume of the probability
space ${\mathbb{P}}_{\!n}$ is
\begin{eqnarray}
V_{n-1}({\mathbb{P}}_{\!n}) \equiv V_{n-1}\left(
({\mathcal{S}}^{n-1})^+\right)= \frac{1}{2^n}\int d\Omega^n =
\frac{ \pi^{n/2}}{2^{n-1} \Gamma\left( \frac{n}{2}\right)}\, .
\end{eqnarray}
\noi Bhattacharyya measure of any set ${\mathcal{A}} \subseteq
({\mathcal{S}}^{n-1})^+$ is then
\begin{eqnarray}
\mu_B({\mathcal{A}}) =
\frac{V_{n-1}({\mathcal{A}})}{V_{n-1}({\mathbb{P}}_{\!n})}\, ,
\label{as6}
\end{eqnarray}
and so particularly the normalization $\mu_B({\mathbb{P}}_{\!n}) =
1$ holds. The reader may see that the Bhattacharyya measure is
indeed a very natural concept. In fact, (\ref{as6}) implies that
the latter is just the Haar measure on ${\mathcal{S}}^{n-1}$. One
could possibly adopt some another (not spherical) metric on the
the probability space $({\mathcal{S}}^{n-1})^+$, but because all
non--singular metric measures are on compact manifolds equivalent
(i.e., they differ only by finite multiplicative functions -
Jacobians) the Bhattacharyya measure will be fully satisfactory
for our purpose. Actually the exclusiveness of Bhattacharyya
measure in non--parametric statistics was already emphasized, for
instance, in~\cite{c10}. Naturalness and simplicity of
Bhattacharyya's measure has been also appreciated in various areas
of physics and engineering ranging from quantum
mechanics~\cite{c11} to statistical pattern recognition and signal
processing~\cite{c12}.

%

\subsubsection{$\boldsymbol{\alpha > 1}$ {\bf{case}}}

\begin{figure}[t]
\vspace{1mm}
\centerline{\hspace{0.6cm}\epsfysize=2.0truein
\epsfbox{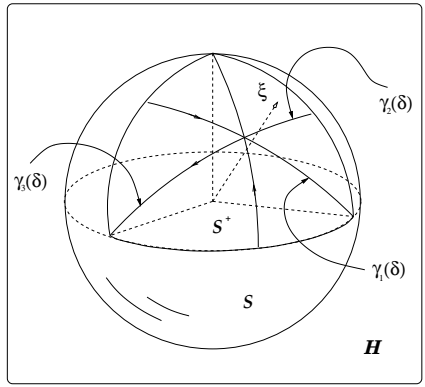}} \vspace{0.5cm}
\caption{\footnotesize{\textit{\textbf{$\!$The family of Lesche's
critical distributions ($\alpha\! >\!1$).} A statistical system
can be represented by points ${\boldsymbol{\xi}}$ on a positive
orthant ${\mathcal{S}}^+$ of the unit sphere ${\mathcal{S}}$ in a
real Hilbert space ${\mathcal{H}}$. 1--parametric families of
Leshe's critical distributions are then represented by arcs
${\gamma}_i(\delta) = \left\{\xi_k(\delta) = \sqrt{\frac{~\delta \
\delta_{ik}}{2} + \left( 1- \frac{\delta}{2} \right) \left(
\frac{1 - \ \delta_{ik}}{n-1} \right)}; \; k \in 1,\ldots, n; \,
\delta \in [0,2] \right\}$. Depicted example corresponds to
${\mathcal{S}} = {\mathcal{S}}^2$.
%
 }}} \label{fig1}
\vspace{2mm}
\end{figure}

Let us now look at the Bhattacharyya measure of the family of
Lesche's critical distributions corresponding to $\alpha > 1$. In
this case the relation (\ref{lesh11}) suggests that the critical
distributions form the $1$--parametric family of distributions
parametrized by $\delta$. Fig.\ref{fig1} indicates that there are
clearly $n$ such families. In contrast to the orthant surface
which has dimension $D=n-1$, the countable set of line--like
$1$--parametric families has the topological dimension $D=1$ and
hence the Bhattacharyya measure of Lesche's critical distributions
is plainly zero.
We wish to ask whether some extension of (\ref{lesh11}) might have
the non--zero measure. We will illustrate now that the answer is
negative. In fact, we will show that with Bhattacharyya measure
approaching $1$ (in the limit of large $n$) all distributions
${\mathcal{P}}\in {\mathbb{P}}_n$ inevitably fulfil Lesche's
condition (\ref{lifs2}). Inasmuch, all distributions which exhibit
the critical behavior encountered in Ref.~\cite{Lesche} have
$\mu_B\rightarrow0$ as $n\rightarrow\infty$. To prove this we
employ the following isoperimetric inequality (also known as
Levy's lemma)~\cite{Mil1}. Let $f$: ${\mathcal{S}}^{n-1} \mapsto
{\mathbb{R}}$ be a $K$--Lipshitz function, i.e., for any pair
${\boldsymbol{\xi}}^{(1)}$, ${\boldsymbol{\xi}}^{(2)} \in
{\mathcal{S}}^{n-1}$
\begin{eqnarray}
||f({\boldsymbol{\xi}}^{(1)}) - f({\boldsymbol{\xi}}^{(2)})|| \leq
K \ ||{\boldsymbol{\xi}}^{(1)} - {\boldsymbol{\xi}}^{(2)}||_2\, .
\end{eqnarray}
Then
\begin{eqnarray}
\frac{V_{n-1}\left( {\boldsymbol{\xi}}\in {\mathcal{S}}^{n-1}; \
\left|f({\boldsymbol{\xi}}) - \int_{{\mathcal{S}}^{n-1}}f d \mu
\right| \ > \ C \right)}{V_{n-1}({\mathcal{S}}^{n-1})} \  \leq \ 4
e^{-\vartheta C^2 n/K}\, , \label{levy1}
\end{eqnarray}
where $\mu$ is the Haar measure on ${\mathcal{S}}^{n-1}$ and
$\vartheta$ is an absolute (i.e., $n$--independent) constant whose
precise form is not important here~\footnote{The metric
$||\ldots||_2$ appearing in the lemma represents the Euclidean
distance inherited from ${\mathbb{R}}^n$ (this is also called the
chordal metric). Note that $||{\boldsymbol{\xi}}^{(1)} -
{\boldsymbol{\xi}}^{(2)} ||_2 = 2\sin(\phi/2) \leq \phi$, with
$\phi$ representing the Bhattacharyya distance.}.

\vspace{3mm}

Let us choose $f({\boldsymbol{\xi}}) =
||{\boldsymbol{\xi}}||_{2\alpha}$. Using the triangle inequality
we have
\begin{eqnarray}
\left|\ ||{\boldsymbol{\xi}}^{(1)}||_{2\alpha} -
||{\boldsymbol{\xi}}^{(2)}||_{2\alpha}     \right| \ \leq \
||{\boldsymbol{\xi}}^{(1)} - {\boldsymbol{\xi}}^{(2)}||_{2\alpha}\
\leq \ ||{\boldsymbol{\xi}}^{(1)} - {\boldsymbol{\xi}}^{(2)}||_2\,
,
\end{eqnarray}
so $||{\boldsymbol{\xi}}||_{2\alpha}$ is $1$--Lipshitz function.
In addition,
\begin{eqnarray}
||{\mathcal{P}} - {\mathcal{Q}}||_1 = \sum_i\left|
({\xi}^{(1)}_i)^2 - ({\xi}^{(2)}_i)^2 \right| = \sum_i \left|\
{\xi}^{(1)}_i - {\xi}^{(2)}_i \right|\left({\xi}^{(1)}_i +
{\xi}^{(2)}_i\right) \geq ||{\boldsymbol{\xi}}^{(1)} -
{\boldsymbol{\xi}}^{(2)} ||^2_2\, . \label{ineq9}
\end{eqnarray}
So particularly when two distributions are $\delta$ close then
their representative points on the sphere fulfil the inequality
\begin{eqnarray}
\left|\ ||{\boldsymbol{\xi}}^{(1)}||_{2\alpha} -
||{\boldsymbol{\xi}}^{(2)}||_{2\alpha}     \right| \ \leq \
\sqrt{\delta}\, . \label{ineq3}
\end{eqnarray}
The next step is to calculate the mean $\int_{{\mathcal{S}}^{n-1}}
f({\boldsymbol{\xi}}) d\mu$. As it stands, this is quite difficult
task but fortunately we may take advantage of the fact that
\begin{eqnarray}
\int_{{\mathcal{S}}^{n-1}}\sum_i^n |\xi_i|^{2\alpha}
d\mu({\boldsymbol{\xi}})\ &=& \ n \int_{{\mathcal{S}}^{n-1}}
|\xi_1|^{2\alpha} d\mu({\boldsymbol{\xi}}) \ = \ \frac{n
\int_0^\pi |\cos(\theta)|^{2\alpha}(\sin(\theta))^{n-2}
d\theta}{\int_{{\mathcal{S}}^{n-1}} (\sin(\theta))^{n-2} d\theta }
\nonumber \\
&=& \ \frac{n \Gamma(n/2)\Gamma(\alpha + 1/2)}{\sqrt{\pi} \
\Gamma(n/2 + \alpha)} \ \sim \ \frac{\Gamma(\alpha + 1/2)
2^\alpha}{\sqrt{\pi}} \ n^{1-\alpha}\, . \label{dis2}
\end{eqnarray}
\vspace{3mm}
(Note that (\ref{dis2}) is true for all  $\alpha >0$. ) Using
Jensen's inequality we then have
\begin{eqnarray}
E\left(||{\boldsymbol{\xi}}||_{2\alpha}  \right) \ \equiv \
\int_{{\mathcal{S}}^{n-1}} ||{\boldsymbol{\xi}}||_{2\alpha} d\mu \
\leq \
\sqrt[2\alpha]{\int_{{\mathcal{S}}^{n-1}}(||{\boldsymbol{\xi}}||_{2\alpha})^{2\alpha}
d\mu} \ = \ \sqrt[2\alpha]{\frac{\Gamma(\alpha + 1/2)
2^\alpha}{\sqrt{\pi}}}\ n^{\frac{1}{2\alpha} -\frac{1}{2}} \, .
%
\label{ineq4}
\end{eqnarray}
On the other hand, because all distributions from ${\mathbb{P}}_n$
fulfill the condition
\begin{equation}
n^{1-\alpha}  \leq  \sum_{i=1}^n p_i^{\alpha}  \leq  1\, , \; \;
\;\; \alpha \geq 1\, ,
\end{equation}
we have that $E\left( ||{\boldsymbol{\xi}}||_{2\alpha} \right)
\geq n^{\frac{1}{2\alpha} - \frac{1}{2}}$. Thus the mean value of
$||{\boldsymbol{\xi}}||_{2\alpha}$ goes to zero as $b
\left(n^{\frac{1}{2\alpha} - \frac{1}{2}}\right)$ where $b=
b(n,\alpha)$ is some bounded function of $n$. Collecting results
(\ref{dis2}) and (\ref{ineq4}) together we can recast Levy's lemma
into form
\begin{eqnarray}
&&\ \mu_B\left( \left| \ ||{\boldsymbol{\xi}}||_{2\alpha} -
E\left( ||{\boldsymbol{\xi}}||_{2\alpha}  \right)
 \right| \,
\leq \, C \right) \geq 1 - 4e^{-\vartheta C^2 n} \nonumber
\\
 \Rightarrow && \ \mu_B \left( \left|
||{\boldsymbol{\xi}}||_{2\alpha} - E\left(
||{\boldsymbol{\xi}}||_{2\alpha}  \right)   \right| \leq \epsilon
\left[E\left( ||{\boldsymbol{\xi}}||_{2\alpha} \right)\right]^p
\right)
 \ \geq \ 1 - 4 \exp\left(- \vartheta \epsilon^2 b^2 n^{\left(1-p\left(
 \frac{\alpha -1}{\alpha} \right)\right)}
 \right)
 \, , \label{ineq5}
\end{eqnarray}
for some $\epsilon > 0$. Note that due to symmetry of
$f({\boldsymbol{\xi}})$ we were allowed to exchanged in
(\ref{levy1}) the averaging over the surface of
${\mathcal{S}}^{n-1}$ for the averaging over the positive octant
$({\mathcal{S}}^{n-1})^+$. Result (\ref{ineq5}) implies that for
any $\epsilon >0$ and any $1< p < \alpha/(\alpha -1)$ the
inequalities
\begin{eqnarray}
&& ||{\boldsymbol{\xi}}||_{2\alpha} \ \geq \
E(||{\boldsymbol{\xi}}||_{2\alpha}) \left( 1 -
\epsilon\left[E\left( ||{\boldsymbol{\xi}}||_{2\alpha}
\right)\right]^{p-1}\right)\ \geq \ E\left(
||{\boldsymbol{\xi}}||_{2\alpha}  \right) e^{-2
\epsilon\left[E\left( ||{\boldsymbol{\xi}}||_{2\alpha}
\right)\right]^{p-1} }\, ,
\nonumber \\
 &&||{\boldsymbol{\xi}}||_{2\alpha}\  \leq \
E(||{\boldsymbol{\xi}}||_{2\alpha})\left( 1 +
\epsilon\left[E\left( ||{\boldsymbol{\xi}}||_{2\alpha}
\right)\right]^{p-1}\right)  \ \leq \ E\left(
||{\boldsymbol{\xi}}||_{2\alpha}  \right) e^{\left[\epsilon
E\left( ||{\boldsymbol{\xi}}||_{2\alpha} \right)\right]^{p-1} }\,
, \label{ineq7}
\end{eqnarray}
hold for almost all ${\boldsymbol{\xi}} \in {\mathbb{P}}_n$ (their
Bhattacharyya measure is arbitrarily close to $1$ as $n$
increases). The fact that ``well behaved" functions are at large
$n$ practically constant on almost entire sphere is known as the
{\em concentration measure phenomenon}~\cite{Mil1,Ball,Gar1}. In
passing, the reader may notice that the relation (\ref{ineq5}) is
a variant of Bernstein--Hoeffding's large deviation
inequality~\cite{Ball,Will}.

\vspace{3mm}

Using now Minkowski's triangle inequality
\begin{eqnarray*}
\left|||{\boldsymbol{\xi}}^{(1)}||_{2\alpha} -
||{\boldsymbol{\xi}}^{(2)}||_{2\alpha} \right| \ \leq \ \left|
||{\boldsymbol{\xi}}^{(1)}||_{2\alpha} -
E(||{\boldsymbol{\xi}}||_{2\alpha})\right| + \left|
||{\boldsymbol{\xi}}^{(2)}||_{2\alpha} -
E(||{\boldsymbol{\xi}}||_{2\alpha}) \right| \ \leq \ 2
\epsilon\left[E\left( ||{\boldsymbol{\xi}}||_{2\alpha}
\right)\right]^p\, ,
\end{eqnarray*}
and bearing in mind (\ref{ineq3}) we can chose $\sqrt{\delta} \geq
2 \epsilon\left[E\left( ||{\boldsymbol{\xi}}||_{2\alpha}
\right)\right]^p$. Consequently (for $n \geq 3$)
\begin{eqnarray}
\frac{\left| {\mathcal{I}}_{\alpha}({\mathcal{P}})
-{\mathcal{I}}_{\alpha}({\mathcal{Q}})
\right|}{{\mathcal{I}}_{\alpha \ {\rm max}}} &=&
\frac{2\alpha}{(\alpha -1) \log_2 n}\left| \log_2\left(
\frac{||{\boldsymbol{\xi}}^{(1)}||_{2\alpha}}{||{\boldsymbol{\xi}}^{(2)}||_{2\alpha}}
\right) \right|\ \leq \ \frac{2\alpha}{(\alpha -1)} \left|\lg
\left( \frac{e^{\epsilon\left[E\left(
||{\boldsymbol{\xi}}||_{2\alpha} \right)\right]^{p-1}
}}{e^{-2\epsilon\left[E\left( ||{\boldsymbol{\xi}}||_{2\alpha}
\right)\right]^{p-1} }}
\right)\right| \nonumber \\
&& \nonumber \\
&=&  \frac{6\alpha \epsilon}{(\alpha -1)}\left[E\left(
||{\boldsymbol{\xi}}||_{2\alpha} \right)\right]^{p-1} \ \leq \
\frac{6\alpha}{(\alpha -1)} \left(
\frac{\delta}{4}\right)^{(p-1)/2p}\, .
\end{eqnarray}
\noi Thus we see that one can always find an appropriate
$\delta_{\varepsilon}$ for every $\varepsilon$, namely
\begin{eqnarray}
\delta_{\varepsilon} \, \leq  \, 4 \left( \frac{\varepsilon
(\alpha -1)}{6\alpha} \right)^{2p/(p-1)}\, ,
\end{eqnarray}
\noi and so the observability condition (\ref{lifs2}) is satisfied
in all cases for which inequalities (\ref{ineq7}) hold.

\subsubsection{$\boldsymbol{0 <\alpha < 1}$ {\bf{case}}}
%
Similar analysis can be performed for critical distributions in
the $\alpha < 1$ case. The corresponding 1--parametric families of
Lesche's critical distributions are represented by arcs
${\varsigma}_i(\delta) = \left\{\xi_k(\delta) =
\sqrt{\left(1-\frac{\delta}{2} \right)\delta_{ik} +
\frac{\delta}{2}\left( \frac{1-\ \delta_{ik}}{n-1} \right)}; \; k
\in \hat{n}; \, \delta \in [0,2] \right\}$. These arcs are
identical with arcs $\gamma_i(\delta)$ depicted in Fig.\ref{fig1}
only the orientation is reversed. Consequently the Bhattacharyya
measure is again zero in this case.

\vspace{3mm}

We may now ask whether there exists some generalization of
(\ref{leshe4}) such that the corresponding measure $\mu_B$ is
non--zero. Answer is again negative. We show now that this is a
consequence of the fact that almost all distributions
${\mathcal{P}}\in {\mathbb{P}}_n$ fulfill Lesche's observability
condition (\ref{lifs2}), while Bhattacharyya's measure of those
distributions which do not comply with the condition (\ref{lifs2})
tends to $0$ at large $n$.

\vspace{3mm}
%
%
To prove this we utilize once again Levy's lemma. In this case we
make identification $f({\boldsymbol{\xi}}) =
||{\boldsymbol{\xi}}^{(2)}||_{2\alpha}/E(
||{\boldsymbol{\xi}}^{(2)}||_{2\alpha} )$. Similarly as in the
previous case we must determine first the asymptotic behavior of
the mean $E\left( ||  {\boldsymbol{\xi}} ||_{2\alpha} \right)$.
This can be achieved by employing Jensen's inequality
\begin{eqnarray}
\sqrt[2\alpha]{\frac{\Gamma(\alpha + 1/2) 2^\alpha}{\sqrt{\pi}}}\
n^{\frac{1}{2\alpha} -\frac{1}{2}} =
\sqrt[2\alpha]{\int_{{\mathcal{S}}^{n-1}}
(||{\boldsymbol{\xi}}||_{2\alpha})^{2\alpha} d\mu} \ \leq \
\int_{{\mathcal{S}}^{n-1}} ||{\boldsymbol{\xi}}||_{2\alpha} d\mu
\, ,
\end{eqnarray}
together with inequality
\begin{equation}
1 \leq \sum_{i=1}^n p_i^{\alpha} \leq n^{1-\alpha}\, , \;\;\; 0 <
\alpha < 1\, .
\end{equation}
Therefore $E\left( || {\boldsymbol{\xi}} ||_{2\alpha} \right)$ is
unbounded at large $n$ and it approaches infinity as
$a(n^{\frac{1}{2\alpha} - \frac{1}{2}} )$ ($a = a(n,\alpha)$ is
some  function with lower and upper bound in $n$). Employing now
the estimate:
\begin{eqnarray}
 \left|\ ||{\boldsymbol{\xi}}^{(1)}||_{2\alpha} -
||{\boldsymbol{\xi}}^{(2)}||_{2\alpha}     \right|
 \leq
||{\boldsymbol{\xi}}^{(1)} - {\boldsymbol{\xi}}^{(2)}||_{2\alpha}
\leq ||{\boldsymbol{\xi}}^{(1)} - {\boldsymbol{\xi}}^{(2)}||_2 \
n^{\frac{1}{2\alpha} - \frac{1}{2}} \leq \sqrt{\delta} \
n^{\frac{1}{2\alpha} - \frac{1}{2}}\, , \label{estimate1}
\end{eqnarray}
(where the triangle and H\"{o}lder inequalities were successively
applied) we obtain that $f({\boldsymbol{\xi}})$ is
$1/\underline{a}$--Lipshitz. Here $\underline{a}$ is the lower
bound\footnote{Clearly $\underline{a} \ \geq \
\sqrt[2\alpha]{\frac{\Gamma(\alpha + 1/2) 2^\alpha}{\sqrt{\pi}}} \
\geq \ 0.529\ldots$.} of $a$.
 Levy's
lemma then implies that
\begin{eqnarray}
\mu_B\left( \left| \frac{||{\boldsymbol{\xi}}||_{2\alpha}}{E\left(
||{\boldsymbol{\xi}}||_{2\alpha} \right)}  -  1 \right| \leq \
\epsilon \right) \ \geq \ 1 - 4 e^{-\vartheta \underline{a}
\epsilon^2 n}\, , \label{ineq6}
\end{eqnarray}
for any $\epsilon >0$. Result (\ref{ineq6}) suggests that for a
sufficiently small $\epsilon$ ($\epsilon \leq 1,59 \ldots$) the
inequality
\begin{eqnarray}
e^{-2\epsilon}  \leq \ 1-\epsilon \ \leq \
\frac{||{\boldsymbol{\xi}}||_{2\alpha}}{E(||{\boldsymbol{\xi}}||_{2\alpha})}
\ \leq \ 1 + \epsilon \ \leq\ e^{\epsilon}\, , \label{ineq8}
\end{eqnarray}
holds for almost all ${\boldsymbol{\xi}} \in {\mathbb{P}}_n$
($\mu_B \rightarrow 1$ as $n\rightarrow\infty$). So we again
encounter the concentration of measure phenomenon - at large $n $
almost all Bhattacharyya measure is concentrated on
${\boldsymbol{\xi}}$'s fulfilling the condition
$||{\boldsymbol{\xi}}||_{2\alpha} \approx
E(||{\boldsymbol{\xi}}||_{2\alpha})$. Using now
\vspace{2mm}
\begin{eqnarray}
\left|\frac{||{\boldsymbol{\xi}}^{(1)}||_{2\alpha}}{E(||{\boldsymbol{\xi}}||_{2\alpha})}
-
\frac{||{\boldsymbol{\xi}}^{(2)}||_{2\alpha}}{E(||{\boldsymbol{\xi}}||_{2\alpha})}
\right| \ \leq \ \left|
\frac{||{\boldsymbol{\xi}}^{(1)}||_{2\alpha}}{E(||{\boldsymbol{\xi}}||_{2\alpha})}
-1 \right| + \left|
\frac{||{\boldsymbol{\xi}}^{(2)}||_{2\alpha}}{E(||{\boldsymbol{\xi}}||_{2\alpha})}
-1 \right| \ \leq \ 2\epsilon \, ,
\end{eqnarray}
\vspace{2mm}
and bearing in mind (\ref{estimate1}) we can set $\delta =
4\epsilon^2 \underline{a}^2$. Consequently (for $n\geq 3$)
\begin{eqnarray}
\frac{\left| {\mathcal{I}}_{\alpha}({\mathcal{P}})
-{\mathcal{I}}_{\alpha}({\mathcal{Q}})
\right|}{{\mathcal{I}}_{\alpha \ {\rm max}}} &=&
\frac{2\alpha}{(1-\alpha) \log_2 n}\left| \log_2\left(
\frac{||{\boldsymbol{\xi}}^{(1)}||_{2\alpha}}{||{\boldsymbol{\xi}}^{(2)}||_{2\alpha}}
\right) \right| \ \leq \ \frac{2\alpha}{(1-\alpha) \lg n}\left|\lg
\left( \frac{e^{\epsilon}}{e^{-2\epsilon}} \right)\right|
\nonumber
\\  & = &\ \frac{6\epsilon \alpha}{(1-\alpha) \lg
n } \ \leq \ \frac{3\sqrt{\delta} \alpha}{(1-\alpha)
\underline{a}}\, .
\end{eqnarray}
\noi As in the previous case we can conclude that it is always
possible to find an appropriate $\delta_{\varepsilon}$ for every
$\varepsilon$, namely
\begin{eqnarray}
\delta_{\varepsilon} \, \leq  \,
\left(\frac{\underline{a}(1-\alpha) \varepsilon }{3\alpha
}\right)^2\, .
\end{eqnarray}
\noi So the observability condition (\ref{lifs2}) is satisfied in
all cases for which (\ref{ineq8}) holds. In passing we should
mention that the underlying reason behind the relations
(\ref{ineq5}) and (\ref{ineq6}) lies in the fact that $n$-spheres
${\mathcal{S}}^n$ equipped with the Bhattacharyya distance
$\phi_n$ and Haar measure $\mu_n$ form the so called {\em normal
Levy family}~\cite{Mil1,Led}. In can be shown~\cite{Mil1} that the
concentration measure phenomenon is an inherent property of any
Levy family.

%
%
%
%

\vspace{3mm}

\vspace{3mm}

The moral of this section can be summarized in the following way:
whenever one selects as the state space for R\'{e}nyi entropies
the space of all discrete statistics then a non--uniform
continuity behavior (i.e., violation of Lesche's condition
(\ref{lifs2})) can be observed for a certain set of distribution
functions in the limit of large $n$. We demonstrated that the
cardinality of such critical distributions is of zero
Bhattacharyya measure in the space of all $n\rightarrow\infty$
probability distributions. One may relate those zero measure
distributions to the so called $l_{\alpha}$--bounded distributions
(i.e., distributions whose $l_\alpha$ norm has non--zero lower
bound for $\alpha >1$ and a finite upper bound for $\alpha<1$).
This can be plainly seen from the fact that for
$l_{\alpha}$--bounded distributions the critical conditions
(\ref{dis2}) and (\ref{ineq8}) cannot be satisfied.


\section{Observability of R\'{e}nyi entropies: Continuous probability
distributions and multifractals}\label{multifrac}

Let us briefly illustrate here that the conditions of absolutely
continuous PDF's or multifractality are themselves sufficiently
restrictive to ensure that the instabilities discussed in the
previous section do not occur. To see this let us consider
Eqs.(\ref{as1}) and (\ref{as2}). The latter imply that for any
${\mathcal{P}}_n$ and ${\mathcal{P}}_n'$ for which the
renormalized R\'{e}nyi entropy exists the following identity holds
\begin{eqnarray}
&~&\nonumber \\
 \frac{|{\mathcal{I}}_{\alpha}({\mathcal{P}}_n) -
{\mathcal{I}}_{\alpha}({\mathcal{P}}_n') |}{{\mathcal{I}}_{\alpha
\; \rm{max}}} &=& \frac{|- D(\alpha)\log_2 l +
{\mathcal{I}}^r_{\alpha}
 + D(\alpha) \log_2 l - {{\mathcal{I}}^r_{\alpha}}' + o(1)
|}{D(\alpha)\log_2 (1/l)}\nonumber \\
&=& \frac{|{\mathcal{I}}^r_{\alpha} -
{{\mathcal{I}}^r_{\alpha}}'|}{D(\alpha) \log_2 (1/l)} + o(1)\, .
\label{frac1}
\end{eqnarray}
\noi Superscript $r$ denotes renormalized quantities. Note
particularly that ${\mathcal{I}}_{\alpha}^r$ are by construction
finite and $n$ (i.e., $l$) independent. Using the fact that $\lg x
\leq (x -1)$ together with H\"{o}lder inequality and
Eq.(\ref{ineq9}) we have for two $\delta$--close distributions
\begin{eqnarray}
|{\mathcal{I}}_{\alpha}({\mathcal{P}}_n) -
{\mathcal{I}}_{\alpha}({\mathcal{P}}_n') | \ \leq \ \frac{2\alpha
k }{|1-\alpha|} \frac{\left|
||{\boldsymbol{\xi}}^{(1)}||_{2\alpha} -
||{\boldsymbol{\xi}}^{(2)}||_{2\alpha}\right|}{\min(||{\boldsymbol{\xi}}||_{2\alpha})}
\ \leq \ \frac{2\alpha k }{|1-\alpha|} \
\frac{\max(||{\boldsymbol{\xi}}||_{2\alpha})}{\min(||{\boldsymbol{\xi}}||_{2\alpha})}
\ \sqrt{\delta}\, ,
\end{eqnarray}
with $k= 1/\lg 2$. Realizing that (\ref{as1}) and (\ref{as2})
imply
\begin{eqnarray}
||{\boldsymbol{\xi}}||_{2\alpha} = e^{ \frac{(1-\alpha)}{2\alpha}
\left(-D(\alpha) \log_2 l  + {\mathcal{I}}^{r}_{\alpha} + o(1)
\right)}\, ,
\end{eqnarray}
we can straightforwardly write that
\begin{eqnarray}
|{\mathcal{I}}^r_{\alpha} - {{\mathcal{I}}^r_{\alpha}}'| + o(1) \
\leq \ \frac{2\alpha k }{|1-\alpha|}\ \sqrt{\delta} \
e^{\frac{(1-\alpha)}{2\alpha}
|({\mathcal{I}}^r_{\alpha})_{\rm{max}} -
({\mathcal{I}}^r_{\alpha})_{\rm{min}}| + o(1)}  \ \equiv  \
\frac{2\alpha k }{|1-\alpha|} \ {\mathcal{B}} \sqrt{\delta}\, .
\label{ex1}
\end{eqnarray}
Here ${\mathcal{B}}$ is an absolute constant representing the
upper bound for the exponential. Gathering results (\ref{frac1})
and (\ref{ex1}) together we can finally write (for $n\geq 2$)
\begin{eqnarray}
\frac{|{\mathcal{I}}_{\alpha}({\mathcal{P}}_n) -
{\mathcal{I}}_{\alpha}({\mathcal{P}}_n') |}{{\mathcal{I}}_{\alpha
\; \rm{max}}} \ \leq \ |{\mathcal{I}}^r_{\alpha} -
{{\mathcal{I}}^r_{\alpha}}'| + o(1) \ \leq \ \frac{2\alpha k
}{|1-\alpha|} \ {\mathcal{B}} \sqrt{\delta}\, .
\end{eqnarray}
It is then clear in this case that one can easily find an
appropriate $\delta_{\varepsilon}$ for every $\varepsilon$, namely
\begin{eqnarray}
\delta_{\varepsilon} \ \leq \ \left( \frac{\varepsilon
(1-\alpha)}{2\alpha k {\mathcal{B}}} \right)^2 \, .
\end{eqnarray}
\noi represents a correct choice. So for all pairs
${\mathcal{P}}_n$ and ${ \mathcal{P}}_n'$ which lead in $n
\longrightarrow \infty$ limit to continuous PDF's (or
multifractals) the Leshe condition (\ref{lifs2}) applies.
%
It is therefore the very definition of systems with absolutely
continuous PDF's/ multifractals (incorporated in Eqs.(\ref{ren6})
and (\ref{mes6})) that naturally avoids the situations with
instability points confronted in the previous section.

%
%
%

\section{Conclusions}\label{remarks}

In this paper we have attempted to make sense of the recent claims
concerning a {\em total} non--observability of R\'{e}nyi's
entropy. We have found that problems have arisen from uncritical
use of Lesche's observability criterion. We have proved that the
latter criterion, as it stands, does not rule out observability of
R\'{e}nyi entropies in large class of systems. Systems with finite
number of microstates or multifractals being examples. This is so
because the structure of the space of distribution functions (or
PDF's) over which such systems operate essentially prohibits an
existence of ``critical" situations considered by Lesche. In cases
where such situations are encountered, namely in systems with
(countable) infinity of microstates, we argue that Lesche's
uniform continuity condition is too tight to serve as a decisive
criterion for the observability.

\vspace{3mm}

In previous works the uniform continuity condition was used to
force observability upon state functions. As we have shown, it is
not just unnecessary to do this but it also causes the Lesche
criterion to produce incorrect results in certain cases. By
identifying the probability distribution with a state variable
this has led to confusion about the observability of R\'{e}nyi's
entropy. Once the uniform continuity condition is dropped, we can
clear up these confusing points. For this purpose we present a
more intuitive concept of observability by allowing the quantity
in question to have a certain amount of ``critical" points
provided that the cardinality of the critical points in
the state space is of zero measure.



\vspace{3mm}

It is definitely interesting to know what the ``critical regions"
correspond to. In case of R\'{e}nyi entropies we offer a partial
reply to this question. Namely, for systems with (countable)
infinity of microstates we show that the critical regions
correspond to the $\delta$--vicinity of $l_{\alpha}$--bounded
distributions. Basically such distributions correspond to
(ultra)rare events which are frequently encountered e.g., in
particle detection (double beta or tritium decays being examples).
We have proved that the Bhattacharyya measure of these
distributions must be zero. As $l_\alpha$--bounded distributions
are not existent in (coarse--grained) multifractals or in systems
with continuous PDF's, neither in systems at thermal equilibrium,
there is no a priori reason to disregard R\'{e}nyi entropies as
observable in the aforementioned instances. On the other hand, it
is known that many systems undergo ``statistics transitions"
(stockmarket bidding and continuous phase transitions with their
exponential--law - power--law distribution ``transitions" may
serve as examples). It might be also expected that in dynamical
systems away from equilibrium transitions to  $l_\alpha$--bounded
statistics may play a relevant r\^{o}le. In any case, one can turn
the sensitivity of R\'{e}nyi entropies to a virtue as it could be
used as a diagnostic instrument for an analysis of
(ultra)rare--event systems, similarly as, for instance,
temperature sensitivity of the susceptibility is used as a
diagnostic tool in continuous phase transitions. We believe that
further investigation in this direction would be of a great value.

\vspace{3mm}

Let us finally stress that there is also a conceptual reason why
the observability \`{a} la Lesche should be viewed with some hint
of scepticism. This is because the observability treated in such a
framework is not a unique concept. Indeed, Lesche's condition can
brand a quantity as observable under one choice of state variables
and as non--observable under a different choice, even if two such
choices overlap in the scope of physical situations they describe.
Typical example is the Gibbs--Shannon entropy. Here, according to
the above criterion, the entropy is observable if the probability
distribution is chosen as the state variable~\cite{Abe2,Lesche}.
On the other hand, if temperature and pressure are state variables
then entropy develops discontinuity in any system which undergoes
first order phase transition (Clausius--Clapeyron equation) and
hence it is not for such systems uniformly continuous function of
state variables, and according to (\ref{lesh1}) (or (\ref{lifs1}))
it is doomed to be non--observable. In this connection it is
interesting to notice that because the parameter $\alpha$ plays
formally r\^{o}le of  inverse temperature~\cite{Sa1,PJ2} one may
expect that various limits may not commute similarly as in
Gibbsian statistical physics. Namely, we may anticipate that
$\lim_{\alpha \rightarrow 1} \lim_{n\rightarrow \infty} \neq
\lim_{n\rightarrow\infty}\lim_{\alpha \rightarrow 1}$. In fact,
Lesche~\cite{Lesche} and other authors~\cite{Abe2} applied the
sequence of limits $\lim_{n\rightarrow\infty}\lim_{\alpha
\rightarrow 1}$. In such a case they concluded that  R\'{e}nyi
entropy of order one (Shannon's entropy) is observable  while the
rest of R\'{e}nyi entropies is not (despite the fact that
R\'{e}nyi entropies are analytic in $\alpha \in {\mathbb{R}}^+$,
see Ref.~\cite{PJ1}). On the other hand, when one utilizes the
``thermodynamical" order, i.e., $\lim_{\alpha \rightarrow 1}
\lim_{n\rightarrow \infty}$, then also R\'{e}nyi's entropy of
order one develops instability points (this may be easily checked
by noticing that unobservability argument presented
in~\cite{Lesche} is continuous in $\alpha =1$). The latter seems
to support our previous comment that Shannon's entropy should not
be uniformly continuous in the space of discrete distribution
functions in order to account, for instance, for the first--order
phase transitions.


\section*{Acknowledgments}

P.J. would like to gratefully acknowledge discussions with S.~Abe,
C.~Isham and D.S.~Brody. P.J. would also like to thank the
Japanese Society for Promotion of Science for financial support.

\section*{References}
%


\begin{thebibliography}{99}
%

\bibitem{Car1} C.~Caratheodory, Math.~Ann. {\bf 67}
(1909) 355; H.A.~Buchdahl, Am.~J.~Phys. {\bf 17} (1949) 212.

\bibitem{Rb1} R.~Balian, {\em From Microphysics to Macrophysics, Methods
and Applications of Statistical Physics, Vol.1} (Springer--Verlag,
Heidelberg, 1991).

\bibitem{Khin2} A.I.~Khinchin, {\em Mathematical Foundations of Statistical Mechanics}
(Dover Publications, Inc., New York, 1949).

\bibitem{Re1} A.~R{\'e}nyi, {\em Probability Theory}
(North-Holland, Amsterdam, 1970); {\em Selected Papers of Alfred
R{\'e}nyi}, Vol.2 (Akad{\'e}mia Kiado, Budapest, 1976).

\bibitem{Pj3} P.~Jizba, cond-mat/0301343.

\bibitem{THC} C.~Tsallis, {\em J. \ Stat. \ Phys.} {\bf 52} (1988) 479;
J.H.~Havrda and F.~Charvat, {\em Kybernatica} {\bf 3} (1967) 30.


\bibitem{Max} http://astrosun.tn.cornell.edu/staff/loredo/bayes/



\bibitem{ABi1} A.~Bialas and W.~Czyz, Acta~Phys.~Pol. B{\bf{31}}
(2000) 2803; Phys. Rev. D{\bf{61}} (2000) 74021; M.K.~Suleymanov,
M.~Sumbera, I.~Zborovsky, hep-ph/0304206 .

\bibitem{CEs1} C.~Essex, C.~Schultzky, A.~Franz and K.H.~Hoffmann,
Physica~A {\bf 284} (2000) 299.



\bibitem{TCH1} T.C.~Halsey, M.H.~Jensen, L.P.~Kadanoff,
I.~Procaccia and B.I.~Shraiman, Phys.~Rev. A~{\bf{33}} (1986)
1141; M.H.~Jensen, L.P.~Kadanoff, A.~Libchaber, I.~Procaccia and
J.~Stavans, Phys.~Rev.~Lett. {\bf{55}} (1985) 2798; K.~Tomita,
H.~Hata, T.~Horita, H.~Mori and T.~Morita, Prog.~Theor.~Phys.
{\bf{80}} (1988) 963; H.G.E.~Hentschel and I.~Procaccia, Physica
D~{\bf{8}} (1983) 435.


\bibitem{Ab1} S.~Abe and Y.~Okamoto (Eds.), {\em Nonextensive Statistical
Mechanics and Its Applications} (Springer--Verlag, New York,
2001).

\bibitem{Lesche} B.~Lesche, J.~Stat.~Phys. {\bf 27} (1982)
419.

\bibitem{Abe2}S.~Abe, Phys.~Rev.E{\bf 66}(2002) 046134


\bibitem{Cac} C.~Cachin, {\it Entropy Measures and Unconditional Security in
Cryptography}, PhD thesis, ETH Zurich, May 1997;
 ftp://ftp.inf.ethz.ch/pub/publications/dissertations/th12187.ps.gz\,
.

\bibitem{Harte1}D.~Harte, {\em Multifractals Theory and
Applications}, (Chapman \& Hall/CRC, New York, 2001);
M.B.~Geilikman, T.V.~Golubeva and V.F.~Pisarenko, Earth \& Plan.
Sci. Let. {\bf 99} (1990) 127.


\bibitem{HM1} H.~Maassen and J.B.M.~Uffink, Phys.~Rev.~Lett.
{\bf{60}} (1988) 1103.

\bibitem{Rudin1} see e.g., W.~Rudin, {\em Functional Analysis},
(McGraw-Hill Companies, New York, 1991).

\bibitem{Ber1} see e.g, F.~Berthier,J.-P.~Diard, L.~Pronzato,
Automatica {\bf 35} (1999) 1605.

\bibitem{Sontag} see e.g., E.D.~Sontag and Y.~Wang, Systems~Control~Lett. {\bf
29}(1997) 279.

\bibitem{PJ1} P.~Jizba and T.~Arimitsu, cond--mat/0207707 .


\bibitem{Cam1} L.L.~Campbell, Information and Control, Vol.8 (1965) 423.


\bibitem{Sa1} H.~Sakaguchi, Prog.~Theor.~Phys. {\bf 81} (1989)
732.


\bibitem{Fed1} J.~Feder, {\it Fractals} (Plenum Press, New York,
1988).

\bibitem{KL1} S.~Kullback and R.~Leibler, Ann.~Math.~Statist. {\bf
22}(1951) 79.

\bibitem{TA1} T.~Arimitsu and N.~Arimitsu, Physica A~{\bf 295}
(2001) 177; J.~Phys.~A: Math.~Gen. {\bf 33} (2000) L235
[CORRIGENDUM: {\bf 34} (2001) 673];  Physica A~{\bf 305} (2002)
218; J.~Phys.: Condens.~Matter {\bf 14} (2002) 2237;
cond-mat/0306042 .

\bibitem{Bha1} A.~Bhattacharyya, Bull.~Calcutta~Math.~Soc. {\bf 35}
(1943) 99; D.C.~Brody and L.P.~Hughston, Proc.~R.~Soc.~London
A{\bf 457} (2001) 1343.

\bibitem{c10} N.Thacker, F.J.Aherne and P.I.Rockett, Kybernetika, {\bf 34} (1997)
363; D.C.Brody and L.P.Hughston, in Disordered and Complex
Systems, (p. 281-288) P.Sollich, A.C.C.Coolen, L.P.Hughston, Eds.
AIP Press, NY (2000); S.S.Dragomir, math.PR/0304240.


\bibitem{c11}W.K.Wootters Phys. Rev. D{\bf 23}  (1981) 357;
H.Araki and G.Raggio, Lett. Math.Phys. {\bf 6} (1982) 237;
D.C.Brody and L.P.Hughston, Phys.Rev.Lett. {\bf 77} (1996) 2851;
D.C.Brody and L.P.Hughston, Proc. R. Soc. Lond. A{\bf 454} (1998)
2445.


\bibitem{c12} K.Kukunaga, Introduction to Statistical pattern Recognition, (Academic Press, Inc., NY, 1990);
R.O.Duda, P.E.Hart, D.G.Stork, Pattern Classification (Wiley,
London, 2000); T.Kailath, IEEE Trans.Comm.Theory {\bf 15} (1967)
52.


\bibitem{Mil1} V.D.~Milman, G.~Schechtman, {\it Asymptotic Theory of Finite Dimensional Normed Spaces}
(Spring--verlag, New York, 1980).


\bibitem{Ball} K.M.~Ball, {\em An elementary introduction to modern convex
geometry}, Flavors of Geometry, Ed. by Silvio Levy, (Cambridge
University Press, New York, 1997).

\bibitem{Gar1} R.J.~Gardner, Bulletin of the American Mathematical
Society, {\bf 39} (2002) 355.

\bibitem{Will} D.~Williams, {\em Probability with martingales} (Cambridge University Press, Cambride,
1991).

\bibitem{Led} M.~Ledoux, {\em Concentration of measure and logarithmic Sobolev
inequalities}, ed. by J.~Az\'{e}ma, M.~\'{E}mery, M.~Ledoux and
M.~Yor, Lecture Notes in Mathematics 1709 (Springer, Berlin,
1999).

\bibitem{PJ2} P.~Jizba and T.~Arimitsu, AIP Conf.~Proc. {\bf 597} (2001)
341.




%












\end{thebibliography}
\end{document}